\documentclass[runningheads]{llncs}
\usepackage{graphicx}
\usepackage{algorithm}
\usepackage[noend]{algpseudocode}
\usepackage{caption} 
\begin{document}
\title{AFDP: An Automated Function Description Prediction Approach to Improve Accuracy of Protein Function Predictions}

\titlerunning{AFDP: An Automated Function Description Prediction Approach}
%
%
\author{Samaneh Jozashoori\inst{1,2}\orcidID{0000-0003-1702-8707}, Amir Jozashoori\inst{3}\orcidID{0000-0002-3922-6774} \and
Heiko Schoof\inst{4}\orcidID{0000-0002-1527-3752}}
\authorrunning{S. Jozashoori et al.}
\institute{L3S Institute, Leibniz University of Hannover, Germany\\
\and
TIB Leibniz Information Centre for Science and Technology, Germany\\
\and
Azad University of Zanjan, Iran\\
\and 
University of Bonn, Germany\\
\email{jozashoori@l3s.de}
\email{}
\email{schoof@uni-bonn.de}}
\maketitle{}              
\begin{abstract}
With the rapid growth in high-throughput biological sequencing technologies and subsequently the amount of produced omics data, it is essential to develop automated methods to annotate the functionality of unknown genes and proteins. There are developed tools such as AHRD applying known proteins characterization to annotate unknown ones. Some other algorithms such as eggNOG apply orthologous groups of proteins to detect the most probable function. However, while the available tools focus on the detection of the most similar characterization, they are not able to generalize and integrate information from multiple homologs while maintaining accuracy.

Here, we devise AFDP, an integrated approach for protein function prediction which benefits from the combination of two available tools, AHRD and eggNOG, to predict the functionality of novel proteins and produce more precise human readable descriptions by applying our stCFExt algorithm. StCFExt creates function descriptions applying available manually curated descriptions in swiss-prot. Using a benchmark dataset we show that the annotations predicted by our approach are more accurate than eggNOG and AHRD annotations.
\end{abstract}
%
%
\begin{figure}[t!]
\includegraphics[width=\textwidth]{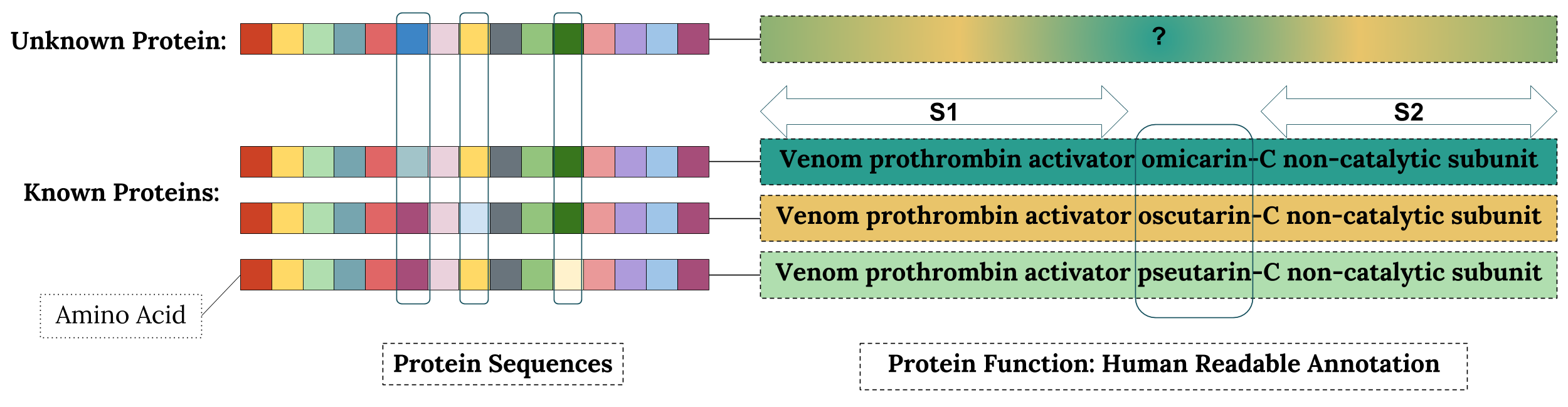}
\caption{\textbf{Example of Representative Description Extraction.} According to the sequence similarities, the unknown protein is predicted to belong to a cluster with three known proteins. The human readable descriptions of these proteins only differ in one token which appears in the middle of the descriptions. Obviously, extracting the LCS only from the whole string will lead to keep part S1 and exclude the S2 which prompts to lose information. This example motivates stCFExt algorithm to repeat the previous steps on remaining prefix and suffix i.e. S2 in this example.   }
\label{figure1}
\end{figure}
\section{Introduction}
Proteins perform a wide variety of important functions within organisms and are necessary for maintaining metabolism and cellular structure. Hence the major problem in understanding the molecular underpinnings of life lies in knowing the functionality of proteins. While molecular experiments provide the most reliable annotation of proteins, their relatively low throughput and restricted scope have led to an increasing role for computational function prediction. Therefore, from the early stages of bioinformatics in the late 80s, the development of high performance and accurate computational tools for predicting the functions of newly identified proteins, was a major focus of the field. 

The functionality of a protein depends on sequence of amino acids from which it has been created and its structure which describes proteins' possible interactions with other molecules. Sequence-based protein function prediction is based on the fact that finding a protein whose function is already characterized experimentally and has significantly similar sequence as the query protein, may reveal some functional aspects of the unknown protein \cite{friedberg2006automated}.

Consequently, based on this similarity principle, methods such as BLAST \cite{altschul1990basic} were developed to compare sequence and structure of proteins and also databases such as Uniprot\footnote{\url{https://www.uniprot.org/}} were developed, which organize function information of proteins and serve as reference to be queried against.

Protein function prediction through similarity searches is based on the evolutionary principles of homology. Generally orthologous genes preserve the same function as their ancestral which makes the identification of them crucial to reliably annotate the functions of proteins that they encode\cite{Reference5}. EggNOG \cite{huerta2015eggnog} is one of the best available tools that provide orthologous groups of proteins and cohesive functional annotation for each.  

After proteins are clustered into the protein families, one representative human readable description needs to be assigned along with the representative sequence to indicate the functionality of the proteins within the group. An automated clustering tool follow one of these approaches to functionally annotate the protein families: \textbf{a.} applying the function annotation that is already assigned to the representative sequence, such as CD-HIT \cite{li2006cd} algorithm, \textbf{b.} generating a function description according to the available annotations of all protein members that are included in the same group, similar to eggNOG approach. Due to the complexities of generating human readable annotation automatically, there are still room to propose an improved algorithm. Therefore, we introduce stCFExt algorithm along with our integrated approach for protein function prediction.

The rest of the paper is structured as follows: in Section 2 the pseudocode and details of stCFExt algorithm is explained while the AFDP approach is presented in Section 3. Section 4 reports the experimental evaluation of stCFExt. Finally, Section 5 presents our conclusions.

\section{The stCFExt Algorithm}
\begin{algorithm}[t!]
\caption{stCFExt’s algorithm}\label{alg:stCFExt}
\begin{algorithmic}[1]
\Require {A text file consists of cluster names and information about all proteins included in each cluster i.e. accession keys and human readable descriptions.}
\Ensure {A text file including cluster names and one representative human readable description for each cluster.}
\For {$i=1$ to $numberOfClusters$}
\State Remove uninformative tokens from descriptions applying defined blacklist.
\State Filter descriptions whose lengths are less than $80\%$ of $maximalLengthOfCluster$.
\State $descriptions$ $\gets$ {extract list of descriptions of the cluster}
\State $st$ $\gets$ MakeSuffixTree($descriptions$)
\State $lcs$ $\gets$ find the longest common substring in $st$ within the length equal to at least $5\%$ of the length of the longest description of the cluster.
\State $prefix\_descs$, $suffix\_descs \gets$ removing $lcs$, extract the former and later remaining parts
\State $pre\_lcs$, $suf\_lcs \gets$ repeat 4 and 5 for both $prefix\_descs$, $suffix\_descs$ lists
\State $pre\_prefix$, $pre\_suffix$, $suf\_prefix$, $suf\_suffix\gets$ repeat 6 for both $pre\_lcs$ and $suf\_lcs$
\State $representativeDescription$ $\gets$ $pre\_prefix + pre\_lcs + pre\_suffix + lcs + suf\_prefix + suf\_lcs + suf\_suffix$
\State \textbf{Write} $reperesentativeDescription$ to the $output$
\EndFor
\State \textbf{return} $output$
\end{algorithmic}
\end{algorithm}
stCFExt\footnote{\url{https://github.com/samiscoding/stCFExt}} or suffix tree based Cluster Function Extraction, is an algorithm we propose for creating representative descriptions for protein families applying Suffix Tree\cite{Reference21}, a fundamental data structure in string processing, and the available human readable descriptions for protein sequences.\\ As it can also be followed in algorithm\ref{alg:stCFExt}, stCFExt initiates with taking all human readable descriptions of the protein sequences included in each protein cluster as input. Considering the general idea of having representative description, the first step of algorithm is filtering and cleaning. In order to provide a generic insight on the functionality of each cluster, a blacklist of tokens which are not informative needs to be defined and be removed from the processing descriptions. Experimentally, we generated the initial blacklist including the following tokens: "putative", "(fragment/ts)", "truncated", "homolog", "probable", and "(predicted)". Another case to be considered is a cluster in which one sequence with totally different description from the others exists. Since such sequences may refer to outliers, stCFExt differentiates between mentioned clusters and those in which more than one description are dissimilar. Accordingly, removing the solitary sequence description is performed in two steps: \textbf{a.} Removal based on the length; the sequence within the description length less that 80\% of the longest description length in the cluster is to be removed. \textbf{b.} Removal based on the informativeness which is to be happened in the later step.\\
After the preliminary filtration, a suffix tree is generated for each cluster and followed by detecting the Longest Common Substring \cite{gusfield1997algorithms} or LCS according to which secondary filtration is performed in order to assure that it is meaningful and informative substring. Afterwards, the LCS detection is performed on each remaining parts of descriptions i.e. prefix and suffix substrings. The motivation and significance of performing later step is shown in Figure\ref{figure1}: as it can be observed in this example, the descriptions of three sequence-based similar proteins to the novel one, only differ in one token which is located in the middle of the whole string and divides it into two common substrings. Subsequently, the remaining prefixes and suffixes are either chosen based on their frequencies in different descriptions or added to the common strings cumulatively. 
\section{The AFDP Approach}
\begin{figure}[t!]
\includegraphics[width=\textwidth]{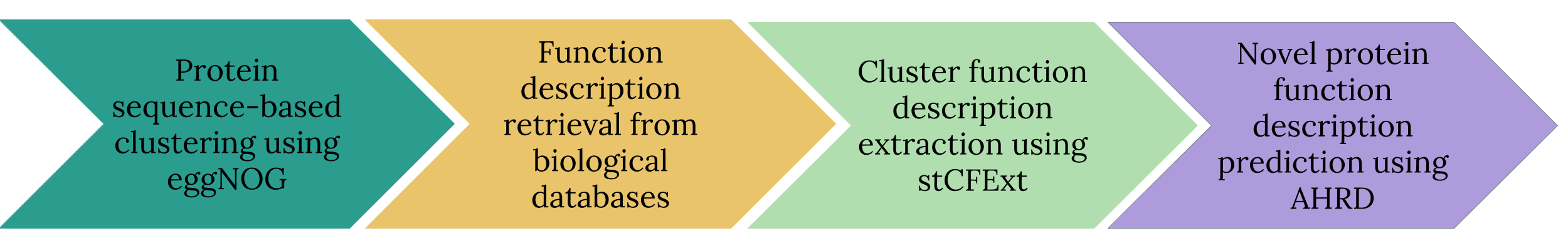}
\caption{\textbf{Automated Function Description Prediction Approach}}
\label{figure2}
\end{figure}
\begin{figure}[t!]
\includegraphics[width=\textwidth]{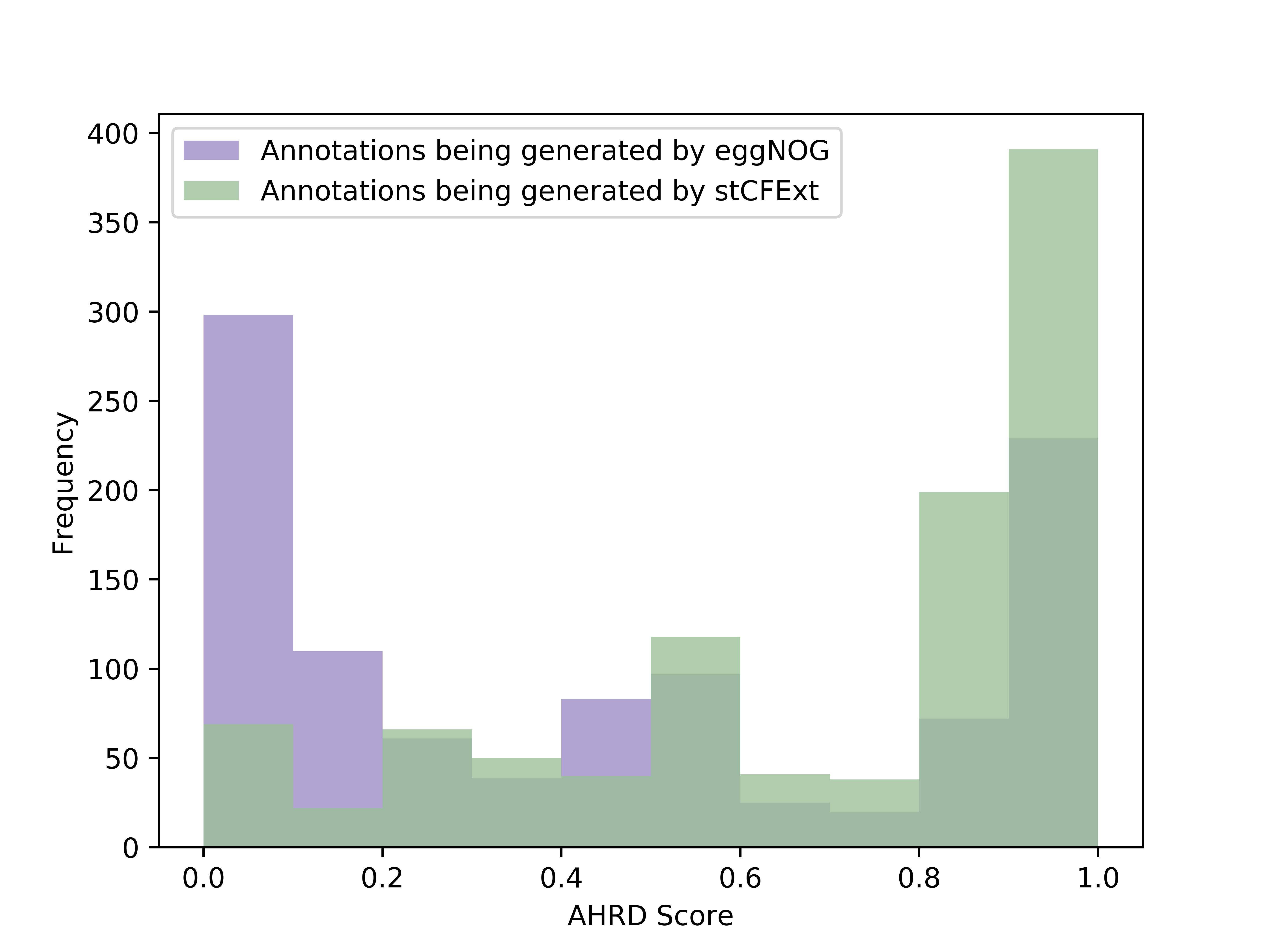}
\caption{\textbf{Comparison between performance of stCFExt and eggNOG}}
\label{figure3}
\end{figure}
AFDP or Automated Function Description Prediction is an integrated approach we devise for protein function prediction. The first 3 steps process the available data about known proteins. As it is shown in Figure\ref{figure2}, the approach starts with clustering proteins into orthologous groups applying eggNOG. Since eggNOG generates new identification numbers and representative descriptions for clustered proteins and does not include the Uniprot accession keys nor descriptions of proteins, an intermediate step is required to retrieve curated descriptions of clustered proteins from Uniprot. Afterwards, an informative representative description is generated for each cluster using stCFExt. This description is annotated to all unknown proteins in the cluster, thus providing a function prediction for all unknown proteins.

\section{Experimental Evaluation}
To evaluate the performance of stCFExt we use the human readable descriptions produced by eggNOG to compare with those generated by stCFExt. In this experiment we address this question: Does stCFExt produce more precise human readable function descriptions? Accordingly, the experiment is designed as follows:\\ 
\indent\textbf{Benchmark:} We use a dataset of protein sequences randomly selected from clusters generated by eggNOG that have manually curated descriptions in swiss-prot (reference description). Their annotations were removed and they were used as input for four predictions: stCFExt, eggNOG, AHRD, and best BLAST result in swiss-prot database\footnote{\url{https://www.uniprot.org/uniprot/?query=reviewed:yes}}.\\ 
\indent\textbf{Metrics:} We evaluate performance based on the number of shared words between prediction and reference description and compute the F Score as harmonic mean of precision and recall \cite{Reference25}. \\
\indent\textbf{Experimental Setup:} We use a version of AHRD \cite{tomato2012tomato} which supports competitor parameters to evaluate different methods simultaneously. Each competitor file includes accession keys of sequences of interest along with their human readable descriptions. Thereby, we expose our algorithm's outcome to be compared with the result of another algorithm in terms of precision and recall.\\             
\indent \textbf{Discussion:}
As it is displayed in the Figure\ref{figure3} and Table\ref{tab1}, the descriptions assigned by stCFExt result in a higher evaluation score than other predictions when compared to swiss-prot curated descriptions, and especially appear more precise than the eggNOG annotations.
\begin{table}[t!]
\label{tab1}
\centering
\begin{tabular}{|c|c|}
\hline
Description creator & evaluation score \\
\hline
stCFExt & 0.4323 \\
\hline
eggNOG & 0.2061 \\
\hline
AHRD & 0.2458\\
\hline
best BLAST hit against swiss-prot DB & 0.4254 \\
\hline
\end{tabular}
\caption{\label{tab1}AHRD evaluation score for a query dataset using reference descriptions generated by different algorithms.}
\end{table}
\section{Conclusion}
We introduced AFDP, an integrated approach for protein function prediction based on stCFExt, an algorithm to produce representative human readable function descriptions. Our evaluation showed that our AFDP approach generates more precise protein annotations compared to eggNOG and other predictions. The key advantage is that high-quality protein orthologous groups from eggNOG can be utilized for human readable function annotation of unknown proteins by generalizing from multiple descriptions.

%
%
%
%
%
%
\bibliographystyle{abbrv}
\bibliography{bibliography.bib}
\end{document}